\documentclass[twocolumn]{aa}
\usepackage{graphicx}
\usepackage{txfonts}
\def \npts {105}
\def \Vrad {14.721}
\def \rms  {9.81}
\def \chs  {1.53}
\def \datt {52250.00 (fixed)}

\def \perB {256.20}

\def \lamB {265.60}
\def \TprB {52175.6}
\def \eccB {0.433}
\def \omgB {161.10}
\def \KapB {564.83}

\def \asnB {11.99}
\def \fmB  {3502}
\def \mssB {17.5}
\def \smaB {0.83}

\def \perC {1296.8}

\def \lamC {31.54}
\def \TprC {51206.4}
\def \eccC {0.284}
\def \omgC {101.83}
\def \KapC {42.71}

\def \asnC {4.88}
\def \fmC  {9.22}
\def \mssC {2.41}
\def \smaC {2.44}
\def \Vradi {14.721}
\def \rmsi  {9.65}
\def \chsi  {1.47}

\def \perBi {255.87}

\def \lamBi {266.23}

\def \eccBi {0.435}
\def \omgBi {161.18}
\def \KapBi {564.75}
\def \IncBi {90.00}

\def \asnBi {11.96}
\def \fmBi  {3487}
\def \mssBi {17.4}
\def \smaBi {0.83}

\def \perCi {1383.4}

\def \lamCi {30.59}

\def \eccCi {0.267}
\def \omgCi {78.99}
\def \KapCi {42.01}
\def \IncCi {90.00}

\def \asnCi {5.15}
\def \fmCi  {9.51}
\def \mssCi {2.44}
\def \smaCi {2.55}
\def \Vral {14.730}
\def \rmsl {23.45}
\def \chil {3.66}

\def \perl {255.86}

\def \laml {263.51}
\def \Tprl {52174.7}
\def \eccl {0.431}
\def \omgl {157.61}
\def \Kapl {573.26}

\def \asnl {12.17}
\def \fml  {3669}
\def \mssl {17.7}
\def \smal {0.83}
\def \linl {4.96}

\def \Vraq {14.752}
\def \rmsq {15.62}
\def \chiq {2.52}

\def \perq {256.04}

\def \lamq {265.04}
\def \Tprq {52175.1}
\def \eccq {0.440}
\def \omgq {159.75}
\def \Kapq {566.75}

\def \asnq {11.98}
\def \fmq  {3497}
\def \mssq {17.5}
\def \smaq {0.83}
\def \linq {10.69}
\def \quaq {-17.29}

\def\abs#1{\left\vert#1\right\vert}
\def \llabel#1{\label{#1}}
\def\figpath{Figures/}
\def\bib#1#2#3#4#5#6#7{\bibitem{#1} {#2} {#3}, {#5}, {#6}, {#7} \par }
\def\bibpp#1#2#3#4#5{\bibitem{#1} {#2} {#3}, {#5} \par }
\def\bibB#1#2#3#4#5#6{\bibitem{#1} {#2} {#3}, {#4} ({#5}, {#6}) \par }
\def\bibBb#1#2#3#4#5#6#7{\bibitem{#1} {#2} {#3}, in {#5} ({#6}) {#7} \par } 

\begin{document}

\titlerunning{A pair of planets around {\small HD}\,202206 or a
  circumbinary planet?}

\title{The CORALIE survey for southern extra-solar planets}

\subtitle{XIII. A pair of planets around {HD}\,202206 or a
  circumbinary planet?}

\author{A.C.M. Correia\inst{1}\fnmsep\inst{2}\fnmsep\inst{3},
           S. Udry\inst{1},
           M. Mayor\inst{1},
	   J. Laskar\inst{3},
           D. Naef\inst{4},
           F. Pepe\inst{1},
           D. Queloz\inst{1} \and
           N.C. Santos\inst{1}\fnmsep\inst{5}
          }

\authorrunning{Correia, Udry, Mayor et al. }

\offprints{A.Correia, 
\email{acorreia@fis.ua.pt}}

\institute{Observatoire de Gen\`eve, 51 ch. des Maillettes, 1290 Sauverny,
           Switzerland
   \and
           Departamento de F\'\i sica da Universidade de Aveiro, Campus
           Universit\'ario de Santiago, 3810-193 Aveiro, Portugal
   \and
           Astronomie et Syst\`emes Dynamiques, IMCCE-CNRS UMR8028,
	   77 Avenue Denfert-Rochereau, 75014 Paris, France
   \and
           European Southern Observatory, Casilla 19001, Santiago 19,
           Chile
   \and
           Centro de Astronomia e Astrof\'\i sica da Universidade de Lisboa,
	   Tapada da Ajuda, 1349-018 Lisboa, Portugal 
              }

\date{\today}

\abstract{ Long-term precise Doppler measurements with the {\small
    CORALIE} spectrograph reveal the presence of a second planet
  orbiting the solar-type star {\small HD}\,202206. The
  radial-velocity combined fit yields companion masses of $m_2\sin
  i=\mssBi\,M_{\rm Jup}$ and $\mssCi\,M_{\rm Jup}$, semi-major axes of
  $a=\smaBi$\,AU and $\smaCi$\,AU, and eccentricities of $e=0.43$
  and $0.27$, respectively. A dynamical analysis of the system
  further shows a 5/1 mean motion resonance between the two planets.
  This system is of particular interest since the inner planet is
  within the brown-dwarf limits while the outer one is much less
  massive.  Therefore, either the inner planet formed simultaneously
  in the protoplanetary disk as a {\it superplanet}, or the outer {\it
    Jupiter-like} planet formed in a circumbinary disk.
  We believe this singular planetary system will provide important
  constraints on planetary formation and migration scenarios.
   
   }

\maketitle

\section{Introduction}
For about 6 years, the {\small CORALIE} planet-search programme in the
southern hemisphere (Udry et al., 2000) has been ongoing on the
1.2 m Euler Swiss telescope, designed, built and operated by the
Geneva Observatory at La Silla Observatory (ESO, Chile).  During
this time, the {\small CORALIE} radial-velocity measurements have allowed
us to detect close to 40 extra-solar planets.  Interestingly,
brown-dwarfs candidates, easier to detect with high-precision Doppler
surveys, seem to be more sparse than exoplanets (Mayor et al.,
1997), especially in the 10-40\,$M_\mathsf{Jup}$ interval (Halbwachs
et al., 2000), the so called {\it brown-dwarf desert}.  Objects
in this domain are very important to understand the brown-dwarf/planet
transition.  The distinction between planets and brown dwarfs may rely
on different considerations such as mass, physics of the interior,
formation mechanism, etc.  From the ``formation'' point of view, the
brown-dwarf companions belong to the low-mass end of the secondaries
formed in binary stars while planets form in the protostellar disk.
Such distinct origins of planetary and multiple-star systems are
clearly emphasized by the two peaks in the observed distribution of
minimum masses of secondaries to solar-type stars (e.g. Udry et al., 2002).  
They strongly suggest different formation and
evolution histories for the two populations: below
10\,$M_\mathsf{Jup}$ the planetary distribution increases with
decreasing mass and is thus not the tail of the stellar binary
distribution.

In this context, the 17.5\,$M_\mathsf{Jup}$ minimum mass companion
detected around {\small HD}\,202206 (Udry et al., 2002,
Paper\,I) provided an interesting massive planet or low-mass
brown-dwarf candidate.  Contrary to {\small HD}\,110833 which was
detected with a comparable $m_2\sin{i}$ companion (Mayor et al.,
1997) and then was shown to be  a stellar binary (Halbwachs
et al., 2000), the distance of {\small HD}\,202206 (46.3 pc)
prevents the {\small HIPPARCOS} astrometric data from constraining the
visual orbit.  At such a distance the expected minimum displacement on
the sky of the star due to the inner companion is only 0.26 mas,
largely insufficient for the {\small HIPPARCOS} precision.  If not due to
unfavorable orbital inclination, the observed low secondary mass sets
the companion close to the limit of the planetary and brown-dwarf
domains.

Apart from the massive planet candidate, the
radial-velocity measurements of {\small HD}\,202206 also revealed an
additional drift with a slope of $\sim$\,43\,ms$^{-1}$yr$^{-1}$
pointing towards the presence of another companion in the system
(Paper\,I).  The long-term follow-up of {\small HD}\,202206 is now
unveiling the nature of the second companion: a planet about ten times
less massive than the inner one. If we assume that the outer planet was formed 
in the stellar protoplanetary disk, the inner planet likely also
formed there, and therefore is not a brown dwarf. This means that
protoplanetary disks may be much more massive than usually thought.
Inversely, if we assume that the inner body was formed as a brown dwarf, then
either the outer planet was also formed as a brown dwarf, or it was formed
in an accretion disk around the binary composed of the main star and the brown
dwarf.

Dynamically, the present system is also very interesting.  The
large mass of the inner planet provokes high perturbations in the
orbit of the outer one.  The system is thus in a very chaotic region, 
but the existence of a 5/1 mean motion  resonance in this region
allows it to stabilize the orbits of the planets in this system.

The stellar properties of {\small HD}\,202206 are briefly recalled in
Sect.\,2, The radial velocities and the new detected companion are
described in Sect.\,3. The stability of the system is examined in
Sect.\,4 and the possible implications of such a system on the
planet versus brown-dwarf formation paradigm are discussed further in
Sect.\,5.


\section{HD\,202206 stellar characteristics}

The {\small HD}\,202206 star was observed by the {\small HIPPARCOS}
astrometric satellite ({\small HIP}\,104903).  
A high-precision 
spectroscopic study of this star was also performed by Santos et
al. (2001) in order to examine the metallicity distribution of
stars hosting planets.  Observed and inferred stellar parameters from
these different sources are summarized in Table\,\ref{T1}, taken
from Paper\,I.

The high metallicity of {\small HD}\,202206 probably accounts for its
over luminosity ($M_V=4.75$, $\sim$\,0.4\,mag brighter than the
expected value for a typical G6 dwarf of solar metallicity) as
$T_\mathsf{eff}$ is also larger than the value expected for a G6
dwarf.

\begin{table}[t!]
\caption{Observed and inferred stellar parameters for {\small HD}\,202206.
Photometric, spectral type and astrometric parameters are from 
{\small HIPPARCOS} (ESA 1997). The atmospheric parameters $T_\mathsf{eff}$, 
log $g$, [Fe/H] are from Santos et al. (2001). The bolometric 
correction is computed from Flower (1996) using the spectroscopic 
$T_\mathsf{eff}$ determination. The given age is derived from the 
Geneva evolutionary models (Schaerer et al., 1993) which also 
provide the mass estimate.}    
\llabel{T1} 
\begin{center}
\begin{tabular}{l l c} \hline \hline
{\bf Parameter}  & $ \quad \quad \quad \quad \quad \quad $ 
& {\bf {\small HD}\,202206}\\
\hline 
Spectral Type & & G6V  \\ 
$ V $   &       & $ 8.08 $  \\ 
$ B-V $ &       & $ 0.714 $  \\ 
$ \pi $ & [mas] & $ 21.58 \pm 1.14 $  \\ 
$ M_V $ &       & $ 4.75 $  \\ 
$ BC $  &       & $ - 0.082 $  \\ 
$ L  $  & $ [L_\odot] $ & $ 1.07 $  \\ 
\hline
 [Fe/H] &       & $ 0.37 \pm 0.07 $  \\ 
$ M $   & $ [M_\odot] $ & $ 1.15 $  \\ 
$ T_\mathsf{eff} $ & [K] & $ 5765 \pm 40 $  \\
 log $ g $ & [cgs] & $ 4.75 \pm 0.20 $  \\
$ v \sin i $ & [km/s] & $ 2.5 $  \\
 age & [Gyr] & $ 5.6 \pm 1.2 $  \\ 
\hline
\end{tabular}
\end{center}
\end{table}

The dispersion of the {\small HIPPARCOS} photometric data of {\small
  HD}\,202206 ($\sigma_\mathsf{Hp}=0.013$ mag) is slightly high for the
star magnitude but some indication of stellar activity is seen in the
spectra.

The radial-velocity jitter associated with intrinsic stellar activity
of rotating solar-type stars may have induced spurious radial-velocity
noise, decreasing our ability to detect planetary low-amplitude
radial-velocity variations.  Although noticeable, the activity level
of {\small HD}\,202206 is not very large (Paper\,I, Fig.\,2). It
adds only some low-level
high-frequency spurious noise in the radial-velocity measurements,
taking into account the small projected rotational velocity of the
star and the long period of the newly detected planet.



\section{Orbital solutions for the HD\,202206 system}
\llabel{orbsolutions}

The {\small CORALIE} observations of {\small HD}\,202206 started in
August 1999.  The obvious variation of the radial velocities allowed
us to announce the detection of a low-mass companion of the star
after one orbital period.  When a second maximum of the
radial-velocity curve was reached, we noticed a slight drift of its
value.  With 95 measurements covering more than 3 orbital periods, a
simultaneous fit of a Keplerian model and a linear drift yielded a
period of 256 days, an eccentricity $e$\,=\,0.43 and a secondary
minimum mass of $17.5\,M_\mathsf{Jup}$ (Paper\,I).  The slope of the
radial-velocity drift was found to be 42.9\,$\mathrm{m
  s}^{-1}\mathrm{yr}^{-1}$, and the available measurements did not allow
us to further constrain the longer-period companion.

After $\npts$ {\small CORALIE} radial-velocity measurements
we are now able to describe the orbit of the third body in the system.
Surprisingly, the former observed drift was not the result of a
stellar companion, but the trace of a not very massive planet in a
$1400$ day orbit with eccentricity $e=0.27$.  Indeed, the outer
planet minimum mass of $\mssCi\,M_\mathsf{Jup}$ is almost ten times
less massive than the inner one.

Using the iterative Levenberg-Marquardt method (Press et al., 1992),
we first attempt to fit the complete set of radial velocities from 
{\small CORALIE} with a single orbiting companion and a linear drift as we did
in Paper\,I (solution {\bf S1}).  This fit implies a companion with $P=255.9$ days,
$e=0.43$ and a minimum mass of $\mssl\,M_\mathsf{Jup}$ (Table\,\ref{T2}),
similar to our previous values (Paper\,I).  
However, the slope of the radial velocity drift now drops to $\linl\,\mathrm{m
s}^{-1}\mathrm{yr}^{-1}$, indicating that something changed after the
consideration of the additional data. Such is also
inadequate, as the velocity residuals exhibit $rms=\rmsl\,\mathrm{m s}^{-1}$,
while the measurement uncertainties are only $\sim$\,8\,$\mathrm{m s}^{-1}$. In
particular, this fit gives a reduced $\sqrt{\chi^2}=\chil$, clearly
casting doubt on the model.  Using a quadratic drift instead of a
linear one (solution {\bf S2}), we get identical values for the companion orbital
parameters (Table\,\ref{T2}), and slightly improve our fit, obtaining
$\sqrt{\chi^2}=\chiq$ (Fig.\,\ref{F1}).

\begin{figure}[t!]
   \centering
    \includegraphics*[height=8.5cm,angle=270]{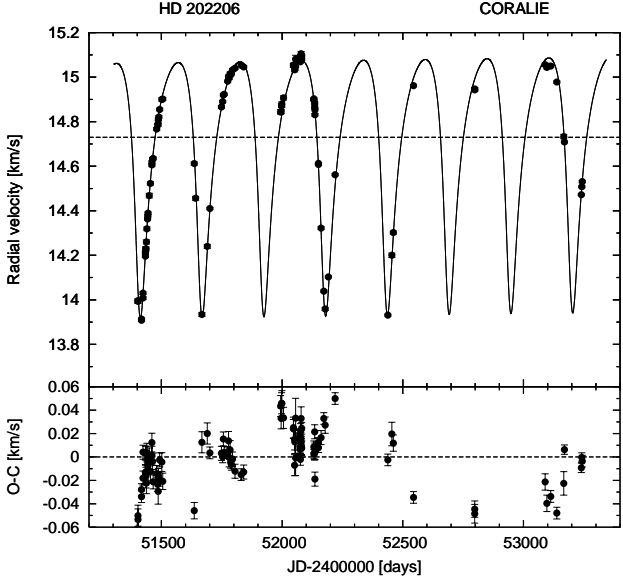}
  \caption{{\small CORALIE} radial velocities for {\small HD}\,202206 
    with a single planet and a linear drift ({\bf S1}). We see that 
    many points lie outside the fitted curve and the value of the reduced
    $\sqrt{\chi^2} = \chil $ is unacceptable.  \llabel{F1} }
\end{figure}

\begin{table}[h]
\caption{Orbital parameters of a single companion orbiting 
{\small HD}\,202206 including drifts in the fit. We consider two
cases: a linear drift ({\bf S1}) and a quadratic drift ({\bf S2}). 
This last model improves 
the fitted solution, but is still unsatisfactory as we also add one 
more degree of freedom. $ \lambda $ is the mean longitude of the date 
($ \lambda = \omega + M $) and errors are given by the standard deviation $
\sigma $.}
\llabel{T2} 
\begin{center}
\begin{tabular}{l l c c} \hline \hline
{\bf Param.}  & {\bf S1 \& S2} & {\bf linear (S1)} & {\bf
quadratic (S2)}   \\ \hline 
rms          & [m/s]                & $ \rmsl $           & $ \rmsq $ \\
$\sqrt{\chi^2}$     &                      & $ \chil $           & $ \chiq $ \\  \hline
Date         & [JD-2400000]         &   \datt             &   \datt   \\ 
$V$          & [km/s]               & $ \Vral \pm 0.001 $ & $ \Vraq \pm 0.001 $ \\  
$P$          & [days]               & $ \perl \pm 0.03  $ & $ \perq \pm 0.03 $ \\ 
$\lambda$    & [deg]                & $ \laml \pm 0.10   $ & $ \lamq \pm 0.12 $ \\ 
$e$          &                      & $ \eccl \pm 0.001 $ & $ \eccq \pm 0.001 $ \\ 
$\omega$     & [deg]                & $ \omgl \pm 0.27   $ & $ \omgq \pm 0.28 $ \\ 
$K$          & [m/s]                & $ \Kapl \pm 1.17  $ & $ \Kapq \pm 1.22 $ \\  
$T$          & [JD-2400000]         & $ \Tprl \pm 0.2   $ & $ \Tprq \pm 0.2 $ \\ \hline
$k_l \; t$   & [m/s/yr]             & $ \linl \pm 0.49  $ & $ \linq \pm 0.55 $ \\
$k_q \; t^2$ & [m/s/yr$^2$]         & $ - $               & $ \quaq \pm 0.66 $ \\ \hline
$a_1 \sin i$ & [$10^{-3}$ AU]       & $ \asnl $           & $ \asnq $ \\
$f (m)$      & [$10^{-9}$ M$_\odot$]& $ \fml  $           & $ \fmq  $ \\
$m_2 \sin i$ & [M$_\mathsf{Jup}$]   & $ \mssl $           & $ \mssq $ \\
$a$          & [AU]                 & $ \smal $           & $ \smaq $ \\ \hline
\end{tabular}
\end{center}
\end{table}

\subsection{Two independent Keplerian fits}

Here we try to fit the radial velocities with two orbiting planetary
companions moving in two elliptical orbits without interaction (solution {\bf S3}).  
The orbits can thus be described by two independent Keplerians as separate
two-body problems, without accounting for mutual planetary
perturbations.

\begin{figure}[t!]
    \includegraphics*[height=8.5cm,angle=270]{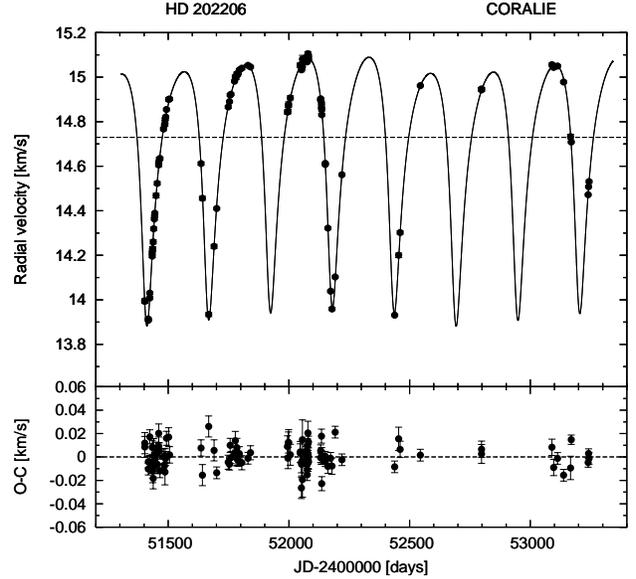} 
  \caption{{\small CORALIE} radial velocities for {\small HD}\,202206 
    with a two independent Keplerian model ({\bf S3}). The residuals 
    are smaller
    than in the case obtained with only one planet (Fig.\,\ref{F1})
    and the value of the reduced $\sqrt{\chi^2}=\chs$ is also better.
  \llabel{F2}}   
\end{figure}

\begin{table}[h!]
\caption{Orbital parameters of two planets orbiting {\small
    HD}\,202206 using a two independent Keplerian model ({\bf S3}). 
    We neglect the gravitational 
interactions between the two planets, but we obtain a better fit than 
using a single planet with a drift (Table\,\ref{T2}). Errors are given 
by the standard deviation $ \sigma $.}
\llabel{T3} 
\begin{center}
\begin{tabular}{l l c c} \hline \hline
{\bf Param.}  & {\bf S3} & {\bf inner} & {\bf
outer}   \\ \hline 
rms          & [m/s]                & \multicolumn{2}{c}{\rms}   \\
$\sqrt{\chi^2}$     &                      & \multicolumn{2}{c}{\chs}   \\ \hline
Date         & [JD-2400000]         & \multicolumn{2}{c}{\datt}  \\ 
$V$          & [km/s]               & \multicolumn{2}{c}{$ \Vrad \pm 0.001 $}  \\  
$P$          & [days]               & $ \perB \pm 0.03  $ & $ \perC \pm 19.1 $ \\ 
$\lambda$    & [deg]                & $ \lamB \pm 0.13  $ & $ \lamC \pm 2.67 $ \\ 
$e$          &                      & $ \eccB \pm 0.001 $ & $ \eccC \pm 0.046 $ \\ 
$\omega$     & [deg]                & $ \omgB \pm 0.31  $ & $ \omgC \pm 6.60 $ \\ 
$K$          & [m/s]                & $ \KapB \pm 1.45  $ & $ \KapC \pm 2.00 $ \\  
$T$          & [JD-2400000]         & $ \TprB \pm 0.2   $ & $ \TprC \pm 29.9 $ \\  \hline
$a_1 \sin i$ & [$10^{-3}$ AU]       & $ \asnB $           & $ \asnC $ \\
$f (m)$      & [$10^{-9}$ M$_\odot$]& $ \fmB  $           & $ \fmC  $ \\
$m_2 \sin i$ & [M$_\mathsf{Jup}$]   & $ \mssB $           & $ \mssC $ \\
$a$          & [AU]                 & $ \smaB $           & $ \smaC $ \\ \hline
\end{tabular}
\end{center}
\end{table}

The two-planet Keplerian fit to the radial velocities using the
Levenberg-Marquardt method yields for the  
inner planet $P$\,=\,$256.2$ days, $e$\,=\,$0.43$ and a minimum mass
of $\mssB\,M_\mathsf{Jup}$, while for the new companion
$P$\,=\,$1297$~days, $e$\,=\,$0.28$ and a minimum mass of
$\mssC\,M_\mathsf{Jup}$ (Table\,\ref{T3}).  The velocity residuals in
this two-planet model drops to $rms$\,=\,$\rms\,\mathrm{m s}^{-1}$ and
the reduced $\sqrt{\chi^2}$ is now $\chs$, clearly suggesting that 
the two companion model represents a significant improvement, even
accounting for the introduction of four additional  free parameters.
The Levenberg-Marquardt minimization method rapidly converges into local minima
of the $ \chi^2 $. However, there is no guarantee that this minimum is global.
Thus, we also fitted our data using a genetic algorithm starting with
arbitrary sets of initial conditions. The found orbital parameters are
identical to the Levenberg-Marquardt solutions. We hence conclude
that our $ \chi^2 $ value is the best for the present data.

The necessity of the second planet is demonstrated visually when we
compare Figs.\,\ref{F1} and \ref{F2} showing {\small CORALIE} radial
velocities and the associated residuals.  
In Fig.\,\ref{F3} we plotted the orbit of the second planet in the radial
velocity residuals of the inner planet.
We also show the frequency analysis of this data and a respective
periodogram of the velocity residuals. The largest frequency peak ($ \approx
0.0007 \, \mathrm{day}^{-1} $) corresponds to the period of the outer planet
and there are no aliases. 
Finally, we computed false alarm probabilities for the second planet through
Monte Carlo simulations by randomly shuffling the data.
The dotted line in the power spectra of the inner planet residuals
(Fig.~\ref{F3}) shows the height of the largest aleatory peak obtained after
100,000 random simulations. The amplitude of this peak is $ 0.023 $~m/s, that
is, about one half of the amplitude of the main peak of the second planet
spectra ($ \approx 0.045 $~m/s).
This gives a false alarm probability of less than $ 10^{-5} $.

\begin{figure}[t!]
   \begin{tabular}{c}
    \includegraphics*[height=8.5cm,angle=270]{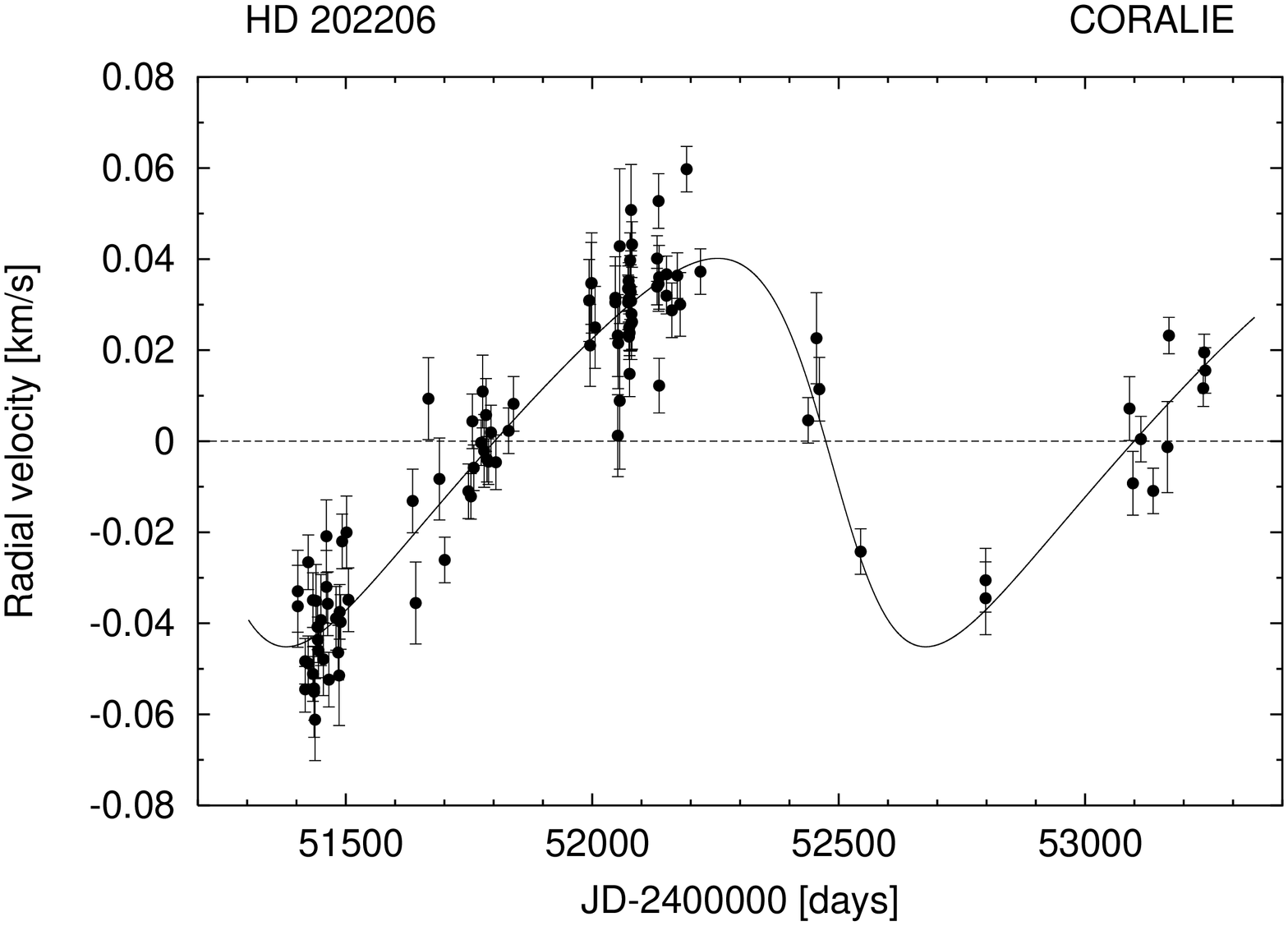} \\
    \includegraphics*[width=8.5cm]{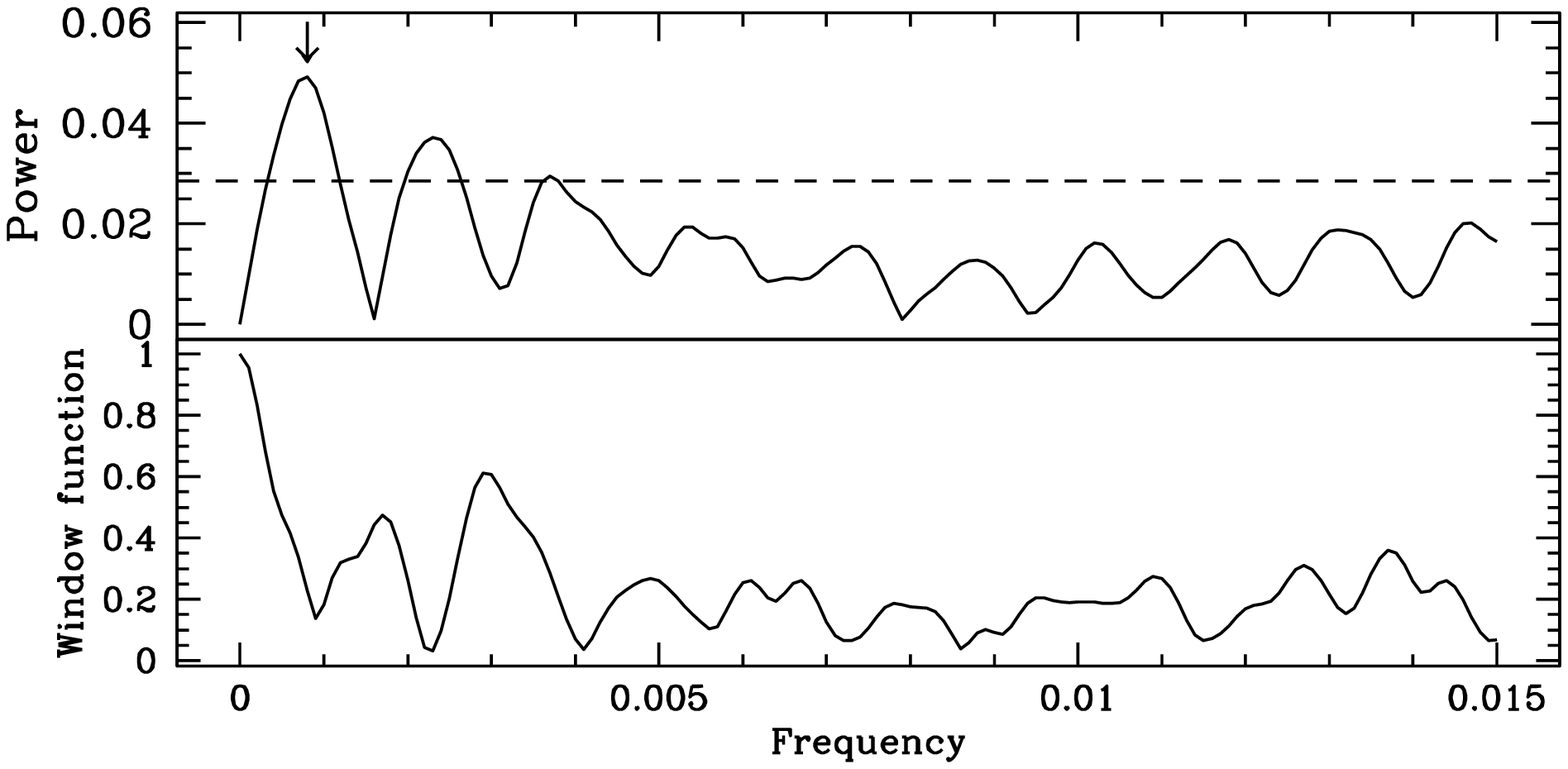}
   \end{tabular}
  \caption{{\small CORALIE} residual radial velocities for 
    {\small HD}\,202206 when the contributions from the inner planet
    are subtracted (top) and respective frequency analysis and periodogram. The
    data acquired before JD=2452200 showed a 
    linear trend that could be provoked by a distant binary companion.
    However, the data acquired after that date clearly shows a short
    period companion. 
    The dotted line in the power spectra shows the height of the largest
    aleatory peak obtained after 100,000 random Monte Carlo simulations.
    This corresponds to a false alarm probability of less than $ 10^{-5} $.
     \llabel{F3}}
\end{figure}

\subsection{Planet-planet interaction}

\llabel{planet_planet_interaction}

Due to the proximity of the two planets and to their high minimum
masses (in particular to the inner planet's huge mass), the
gravitational interactions between these two bodies will be quite
strong.
This prompt us to fit the observational data using a 3-body model (solution {\bf S4}),
similarly to what has been done for the system GJ\,876 (Laughlin and
Chambers, 2001, Laughlin et al., 2004).  Assuming co-planar
motion perpendicular to the plane of the sky, we get slightly better
results for $\sqrt{\chi^2}$ and velocity residuals (Table\,\ref{T4})
than we got for the two-Keplerian fit.  The improvement in our fit is
not significant, but there is a striking difference: the 3-body fitted
orbital parameters of the outer planet show important deviations from
the two-Keplerian case.
We then conclude that, although we still cannot detect the
planet-planet interaction in the present data, we will soon be able to
do so.  We have been following the {\small HD}\,202206 system for
about five years and we expect to see this gravitational interaction
in less than another five years. Thus, two complete orbital
revolutions of the outer planet around the star should be enough.  In
Fig.\,\ref{F4} we plot the two fitting models evolving in time and we
clearly see detectable deviations between the two cruves appearing in 
a near future.  

Finally,
we also fitted the data with a 3-body model where the inclination of
the orbital planes was free to vary (as well as the node of the outer
planet).  We were unable to improve our
fit, even though we have increased the number of free parameters by
three.  Therefore, the inclination of the planets remains unknown, as
do their real masses.

\begin{figure}[t]
   \begin{tabular}{c}
    \includegraphics*[height=8.5cm,angle=270]{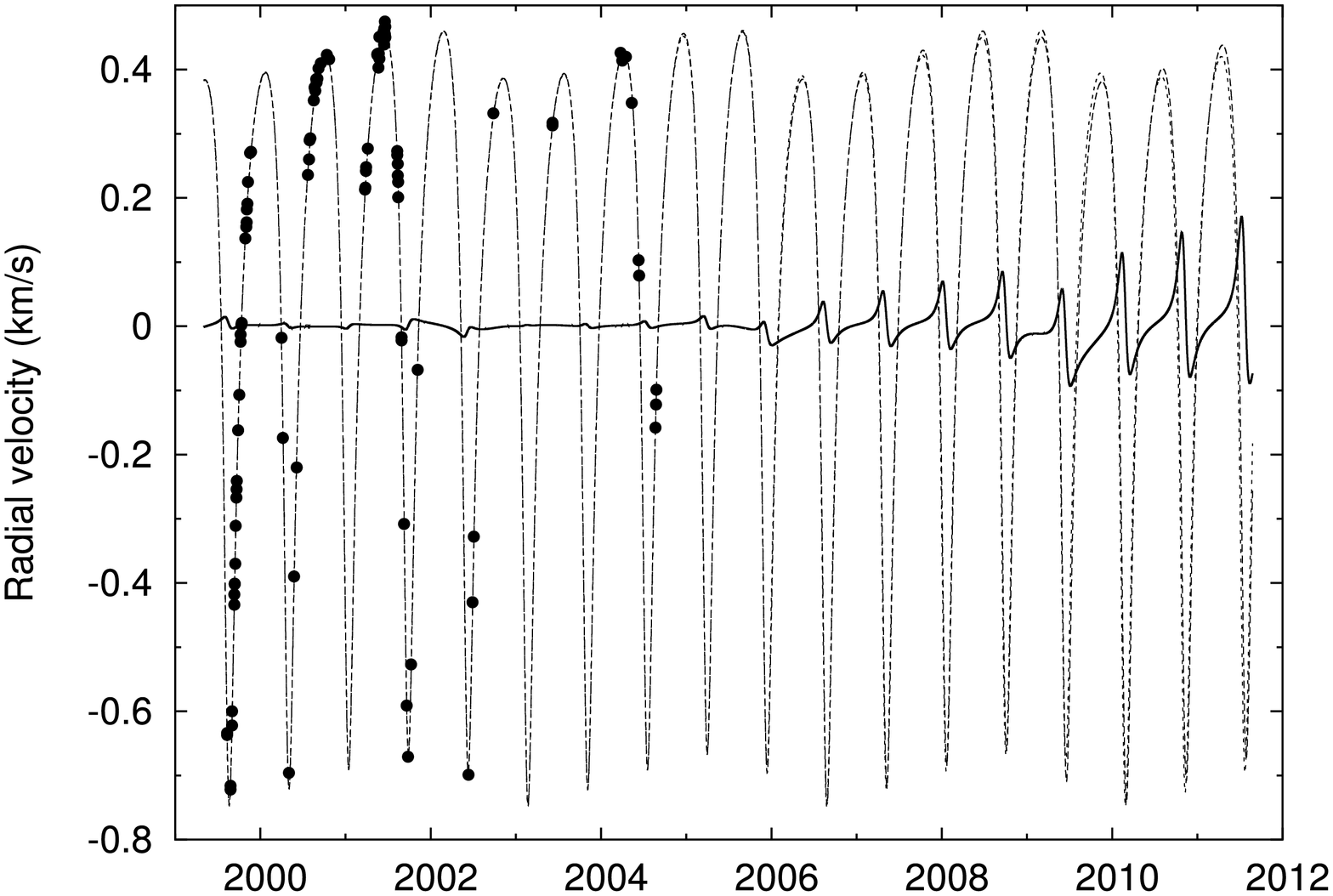} \\
    \includegraphics*[height=8.5cm,angle=270]{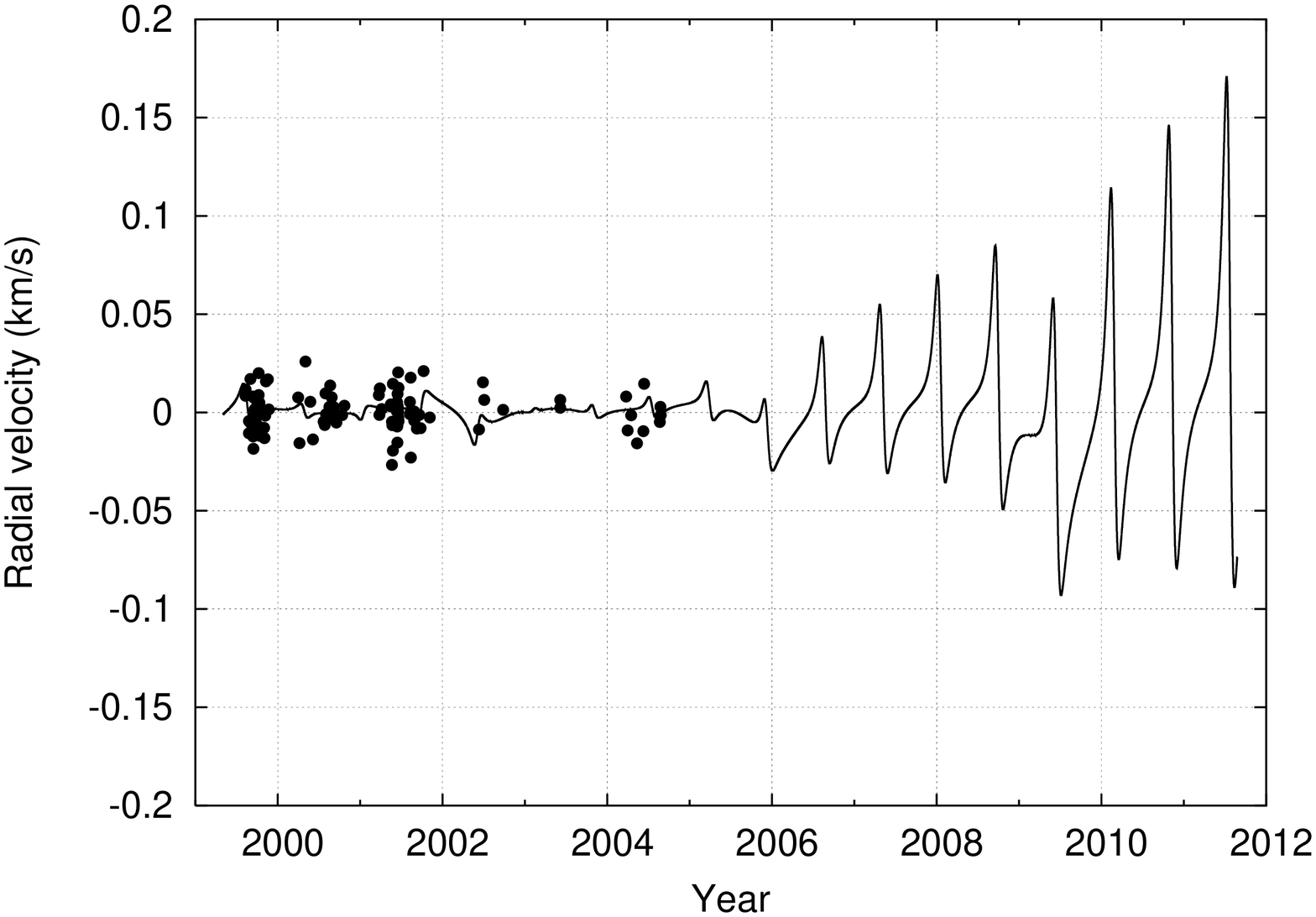}
   \end{tabular}
  \caption{Radial velocities differences between the two independent 
    Keplerian model (Table\,\ref{T3}) and the 3-body model
    (Table\,\ref{T4}).  Data coincides at JD=2452250 (Dec. 6th 2001).
    In the bottom figure we plotted the velocity residuals of the two Keplerian fit.
    Since {\small CORALIE}'s precision is about 8~m/s for this star, we expect
    to observe these differences in the coming years.  \llabel{F4}}
\end{figure}

\begin{table}
\caption{Orbital parameters of two planets orbiting {\small HD} 202206 
using 3-body model ({\bf S4}). We take into account the gravitational
interactions between the two planets, but we obtain a similar fit to
the two-Keplerian model (Table\,\ref{T3}). However, the orbital
parameters of the outer planet are different.
Errors are given by the standard deviation $ \sigma $.}
\llabel{T4} 
\begin{center}
\begin{tabular}{l l c c} \hline \hline
{\bf Param.}  & {\bf S4} & {\bf inner} & {\bf
outer}   \\ \hline 
rms          & [m/s]                & \multicolumn{2}{c}{\rmsi}   \\
$\sqrt{\chi^2}$     &                      & \multicolumn{2}{c}{\chsi}   \\ \hline
Date         & [JD-2400000]         & \multicolumn{2}{c}{\datt}   \\ 
$V$          & [km/s]               & \multicolumn{2}{c}{$ \Vradi \pm 0.001 $}  \\  
$P$          & [days]               & $ \perBi \pm 0.06  $ & $ \perCi \pm 18.4 $ \\ 
$\lambda$    & [deg]                & $ \lamBi \pm 0.18  $ & $ \lamCi \pm 2.84 $ \\ 
$e$          &                      & $ \eccBi \pm 0.001 $ & $ \eccCi \pm 0.021 $ \\ 
$\omega$     & [deg]                & $ \omgBi \pm 0.30  $ & $ \omgCi \pm 6.65 $ \\ 
$K$          & [m/s]                & $ \KapBi \pm 1.34  $ & $ \KapCi \pm 1.50 $ \\  
$i$          & [deg]                & $ \IncBi $ (fixed)   & $ \IncCi $ (fixed) \\  \hline
$a_1 \sin i$ & [$10^{-3}$ AU]       & $ \asnBi $           & $ \asnCi $ \\
$f (m)$      & [$10^{-9}$ M$_\odot$]& $ \fmBi  $           & $ \fmCi  $ \\
$m_2 \sin i$ & [M$_\mathsf{Jup}$]   & $ \mssBi $           & $ \mssCi $ \\
$a$          & [AU]                 & $ \smaBi $           & $ \smaCi $ \\ \hline
\end{tabular}
\end{center}
\end{table}
  

\section{Orbital stability}
\llabel{section_orbital_stability}

In this section we briefly analyze the dynamical stability of the
orbital parameters obtained in the previous section.  A more detailed
study of the system behavior will be presented in a forthcoming paper.

\begin{figure*}[t!]
   \begin{tabular}{c c}
     two independent Keplerian model ({\bf S3}) & 
     3-body dynamical model ({\bf S4}) \\
    \includegraphics*[width=8.5cm]{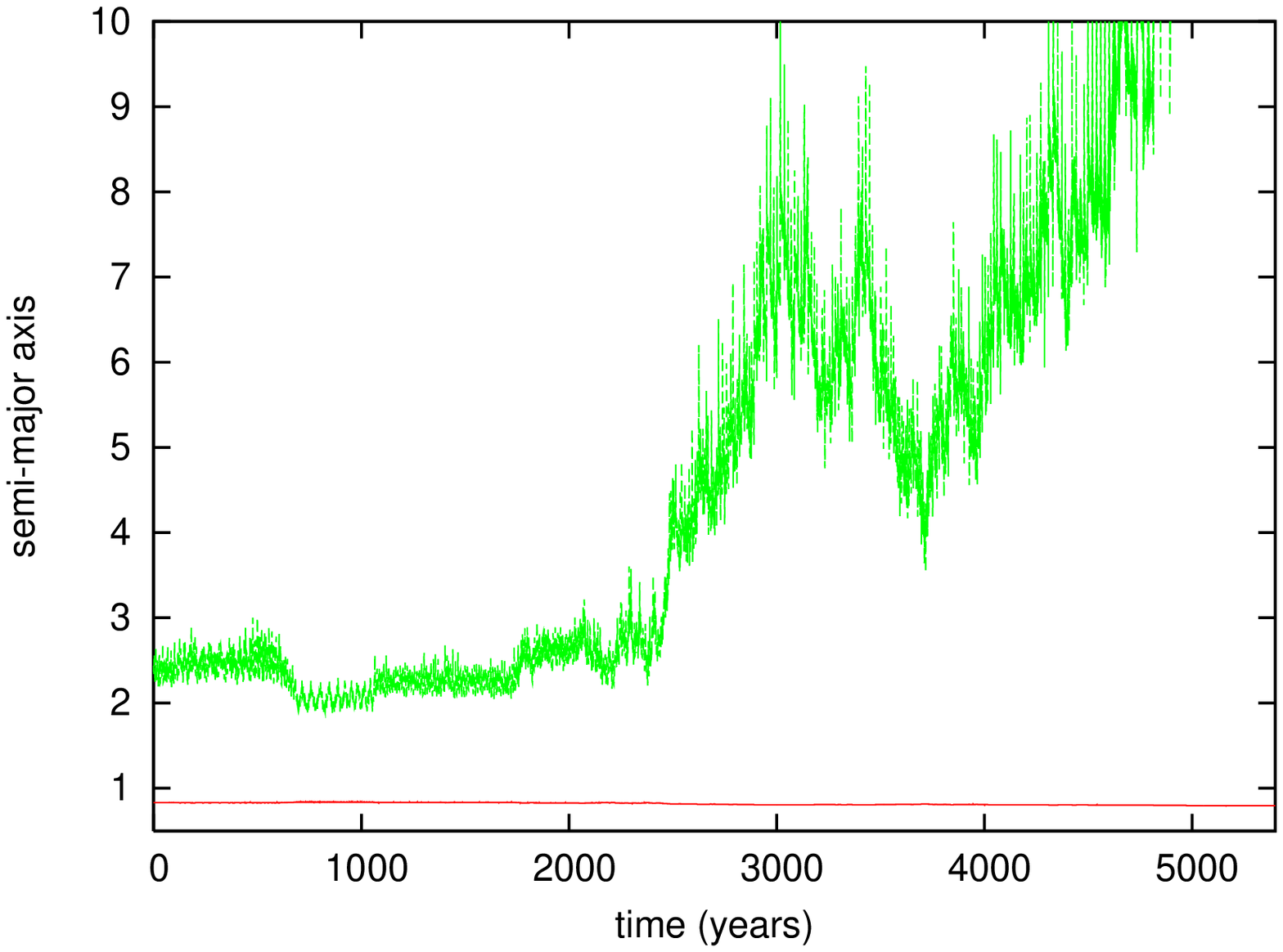} &
    \includegraphics*[width=8.5cm]{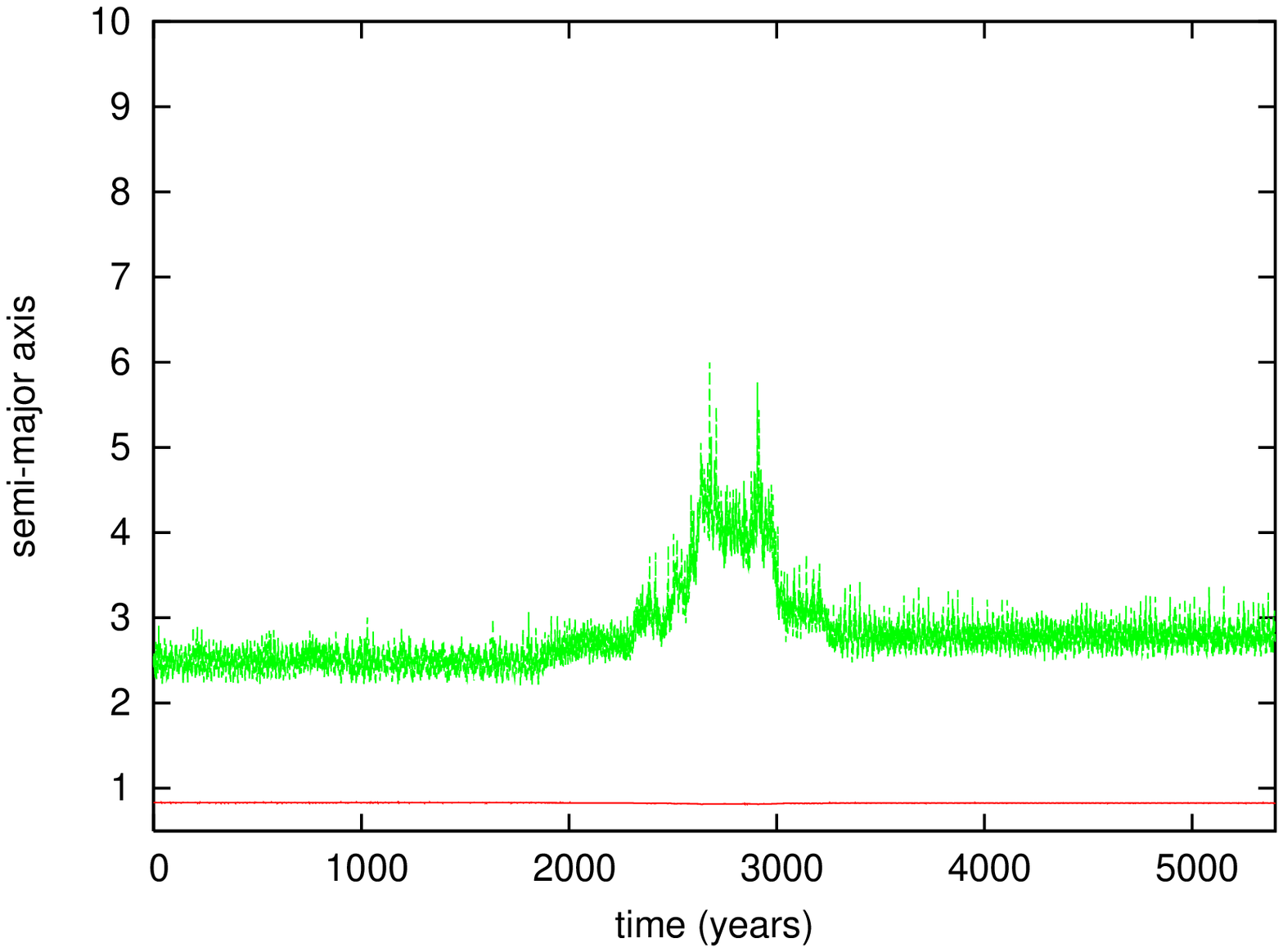} \\
    \includegraphics*[width=8.5cm]{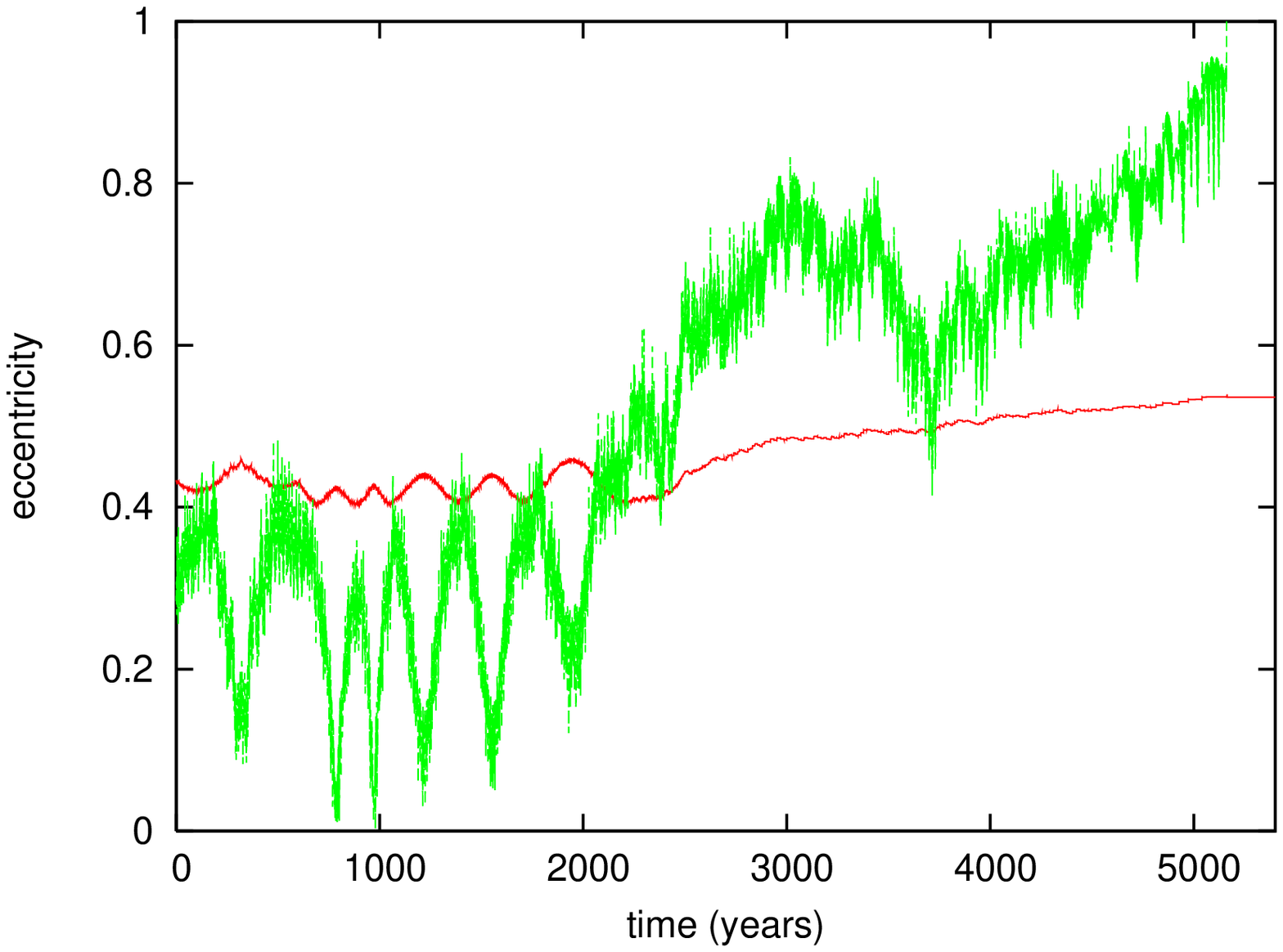} &
    \includegraphics*[width=8.5cm]{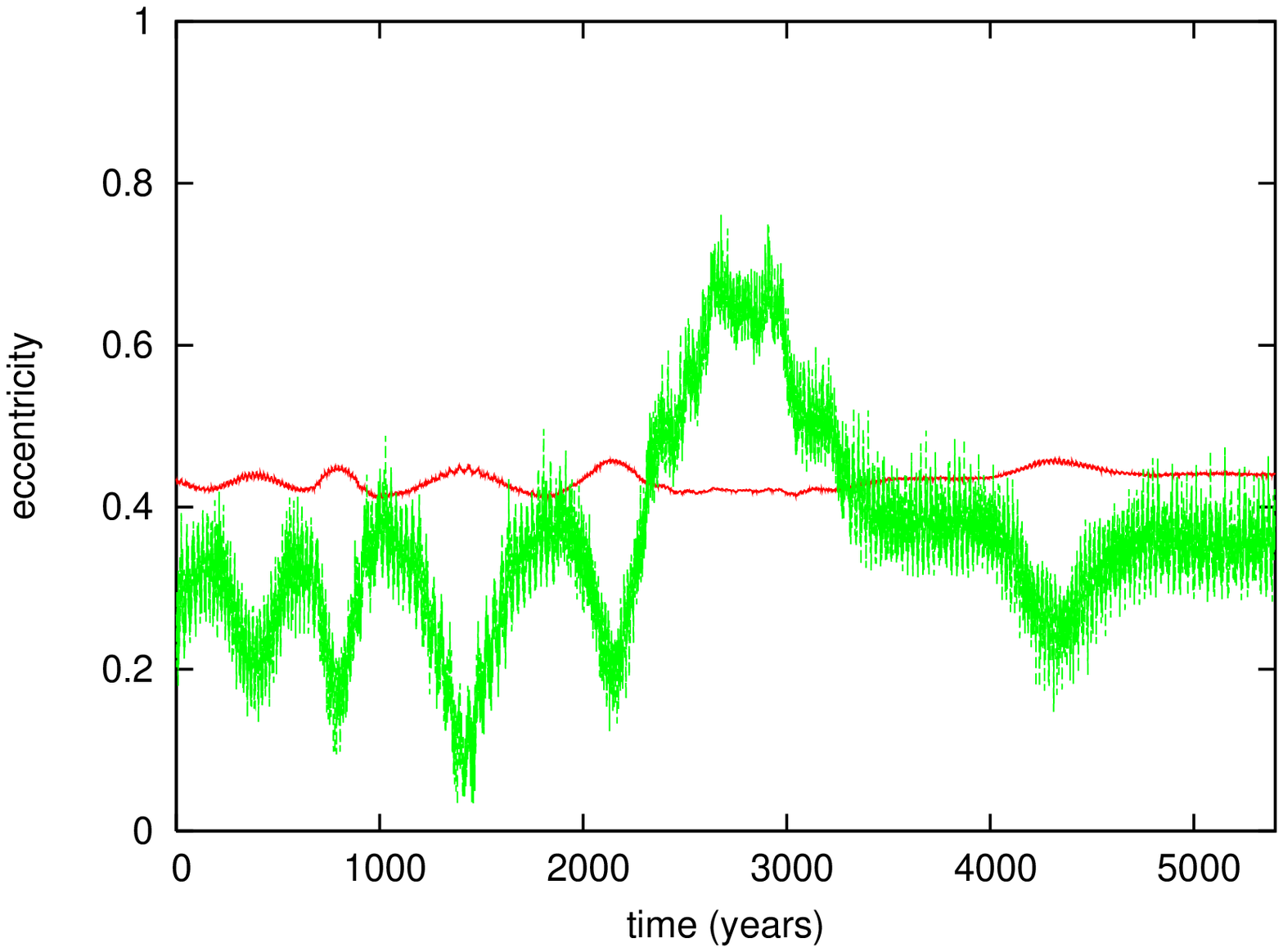} 
   \end{tabular}
  \caption{Dynamical evolution of the semi-major axes and eccentricity for
    two different sets of initial parameters. On the left we plotted
    the evolution for S3 initial parameters obtained with a two
    independent Keplerian model (Table\,\ref{T3}), while on the right
    we used the S4 initial parameters from the 3-body dynamical model
    (Table\,\ref{T4}). Both sets of initial parameters are unstable,
    although S4 is a little better (the outer planet
    is only lost after forty thousand years).  \llabel{F5}}
\end{figure*}

\subsection{Dynamical evolution}

In last section we saw that there were two different models to fit the
observational data: a simplified model using independent Keplerian
orbits for each planet ({\bf S3}) and a 3-body dynamical model
({\bf S4}).  Tracking the dynamical evolution of both sets of
parameters in the future, we find that the two systems become unstable
in a few thousand years (Fig.\,\ref{F5}).  For the initial parameters
obtained with the orbital solution S3, the outer planet is lost after
only five thousand years, the same happening with the system S4 at
about forty thousand years.  This last solution is a slightly better 
determination of the planetary system around {\small HD}\,202206, although 
it is still very unsatisfactory.  It can nevertheless be used as a starting
point for a dynamical study of this system.

\subsection{Stable solutions}
\label{sec4.2}

Since the estimated age of the {\small HD}\,202206 star is about
5\,Gyr (Table\,\ref{T1}), it is  clear that the previous orbits are 
not good.  
One reason is that the fitted parameters 
still present some uncertainties around the best fitted value.  
This is particularly true
for the outer planet, with a small semi-amplitude variation of about
40~m/s.  Moreover, in order to fit our observational data to the
theoretical radial-velocity curve, we used the iterative
Levenberg-Marquardt method.  This method converges to a
minimum  $\chi^2$, but other close local minima may represent as
well a good fit for our data.  Additionally, there may exist additional 
planets in the system that will also perturb the present solution.
We should thus consider that the set of parameters given in
section~\ref{orbsolutions}  constitutes the best determination 
one can do so far, and we will  search 
  for  more stable solutions in its  vicinity.

Starting with the orbital solution S4, obtained with the 3-body model
(Table\,\ref{T4}), we have searched  for possible nearby stable zones.  
Since the orbit of the inner planet is well established, 
with small standard errors,
we have kept the parameters of this planet constant.
We also did not change the inclination of the orbital planes, keeping
both at $90^\circ$.  For the outer planet we let $a$, $\lambda$,
$e$ and $\omega$ vary.  Typically, as in Fig.\,\ref{F6}
we have fixed $e$ and $\lambda$ to 
specific values, and have spanned the $(a,\omega)$ plane of initial conditions
with a step size of  $0.005$ AU for $a$ and 1 degree for $\omega$.
For each initial condition, the orbit of the planets are integrated over 
2000 years with  the  symplectic integrator SABAC4 of 
Laskar and Robutel (2001), using a step size of $0.02$ year.
The stability of the orbit is then measured by 
frequency analysis (Laskar, 1990, 1993). Practically, a refined 
determination of the mean motion $n_2, n_2'$ of the outer planet 
is obtained over two consecutive time interval of length $T=1000$ years, 
and the measure of  the difference $D = \abs{n_2-n_2'}/T$ 
(in deg/yr${}^2$ in Fig.\,\ref{F6})
is a measure of the chaotic diffusion of the trajectory. 
It should be close to zero for a regular solution and 
high values will correspond to strong chaotic motion
(see Laskar, 1993 for more details).

In the present case a regular motion will require  $ D < 10^{-6}$.
We find that the vicinity of the {\small HD}\,202206 system is very
chaotic (light grey region of Fig.\,\ref{F6}) 
and the majority of the initial conditions will rapidly become
unstable.  Because of the two planets' proximity and large values of the  masses
and eccentricities, the chaotic behavior was expected. 
We nevertheless find a small  region of initial conditions
(the darker region of  Fig.\,\ref{F6})
with very small diffusion and where the trajectories 
 remain stable for several million years.  
 These orbital solutions  correspond to the resonant 
 island of  an orbital  5/1 mean motion resonance.

\begin{figure}[t!]
   \begin{tabular}{c}
    \includegraphics*[width=8.5cm]{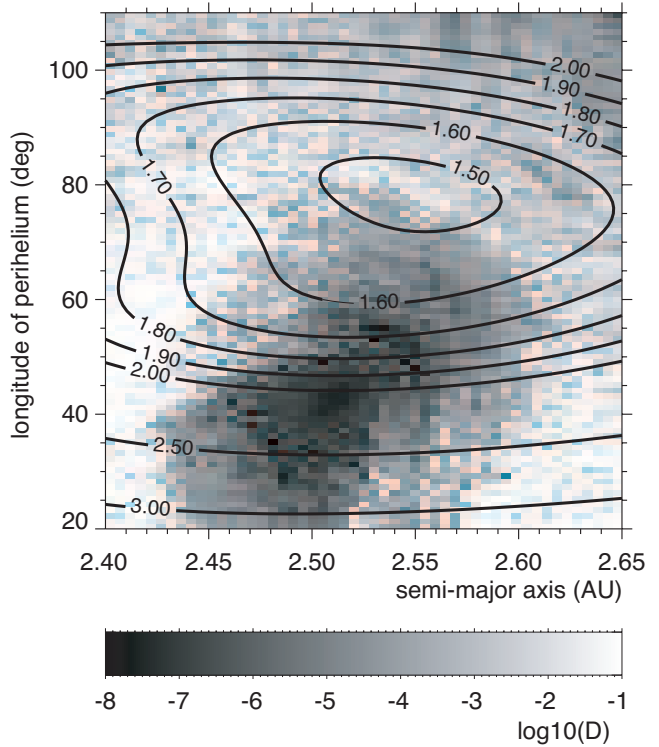}
   \end{tabular}
  \caption{Global view of the dynamics of the {\small HD}\,202206 
    system for variations of the perihelium and semi-major axes of the
    outer planet. Light grey areas correspond to high orbital diffusion
    (instability) and dark areas to low diffusion (stable orbits).
    The grey scale is the stability index ($D$) obtained through a frequency 
    analysis of the longitude of the outer planet over two 
    consecutive time intervals of 1000 yr.
    Labeled lines give the value of $ \sqrt{\chi^2} $ obtained for
    each choice of parameters. Initial conditions in the dark spot
    stable zone (with $\log_{10}(D) < -6$)
    are trapped in a 5/1 mean motion resonance.
    \llabel{F6}}
\end{figure}

Labeled lines of Figure~\ref{F6} give the value
of $\sqrt{\chi^2}$ obtained for each choice of parameters.
We observe that the minimum $\chi^2$  obtained for the present data is
effectively in a zone of high orbital diffusion.  Stable orbits can
only be found inside the dark spot, which corresponds to the 5/1 mean
motion resonance.  
In order to find stable solutions
coherent with our data, we need to increase $\chi^2$ until we get
initial conditions inside this resonant zone.  Thus, the best fit that
provides a stable orbital solution will present $\sqrt{\chi^2} \sim
1.7$, which is still acceptable.  For instance, choosing
$\omega=55.50^\circ$ and $a=2.542$\,AU (solution {\bf S5}), 
we have $\sqrt{\chi^2}=1.67$ (Table\,\ref{T5}).

\begin{table}[th!]
\caption{Stable orbital parameters for the two planets orbiting
{\small HD}\,202206. Using the orbital solution S4
(Table\,\ref{T4}), we chose the values of the perihelium and the
semi-major axes of the outer planet such that the system becomes 
stable. The new system is in a $5/1$ mean motion resonance.}
\llabel{T5} 
\begin{center}
\begin{tabular}{l l c c} \hline \hline
{\bf Param.}  & $ {\bf S5} $ & {\bf inner} & {\bf
outer}   \\ \hline 
 $a$          & [AU]                 & $ 0.83040 $ & $ {\bf 2.54200} $ \\ 
 $\lambda$    & [deg]                & $ 266.22864 $ & $ 30.58643 $ \\ 
 $e$          &                      & $ 0.43492 $ & $ 0.26692 $ \\ 
 $\omega$     & [deg]                & $ 161.18256 $ & $ {\bf 55.50000 } $ \\ 
 $i$          & [deg]                & $ 90.00000 $ & $ 90.00000 $ \\  
 $m $       & [M$_\mathsf{Jup}$]     & $ 17.42774 $ & $ 2.43653 $ \\ \hline
Date         & [JD-2400000]         & \multicolumn{2}{c}{52250.00}   \\ 
rms          & [m/s]                & \multicolumn{2}{c}{\bf 10.73}   \\
$\sqrt{\chi^2}$     &                      & \multicolumn{2}{c}{\bf 1.67}   \\ \hline
\end{tabular}
\end{center}
\end{table}

\begin{figure}[h]
   \begin{tabular}{c}
    \includegraphics*[width=8.5cm]{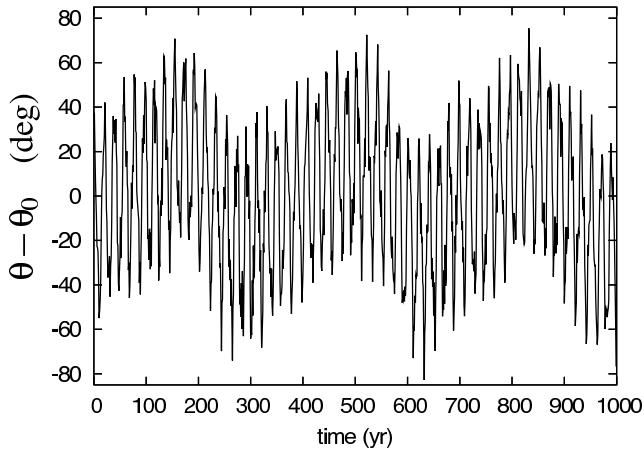}
   \end{tabular}
  \caption{In the orbital solution S5 (Table\,\ref{T5}),
  the resonant argument $\theta = \lambda_1 - 5 \lambda_2 +  g_1 \, t + 3 g_2 \, t$
  is in libration around $\theta_0 = 76.914$ deg, with a libration 
period   $P_{\theta} \approx 19.4$ yr, and an amplitude of about $37$ degrees.
    \llabel{F7}}
\end{figure}

Henceforward, we will consider that  the solution S5
(with orbital parameters given in Table\,\ref{T5})
is more representative of the  real behavior of  
the {\small HD}\,202206 planetary system.
Ideally, we would like that the best fit to the observation 
would also be in a stable region, but we assume that in the present case, 
this requirement is not satisfied because of the limited 
time span and resolution of the observations that do not 
allow us to solve precisely for the outer planet elements. 
In particular, we have not been able yet to solve for the
mutual
inclination of the planets that may also shift the location of the regular 
resonant island.


For the orbital solution S5, the main resonant argument is 
\begin{equation}
\theta = \lambda_1 - 5 \lambda_2 +  g_1 \, t + 3 g_2 \, t
\end{equation}
where $g_1$ and $g_2$ are fundamental secular frequencies of the
system related to the perihelion of the inner and outer planet
respectively (see Laskar, 1990). Both are retrograde,  with periods 
$P_{g_1} \approx 399000$ yr and $P_{g_2} \approx 339$ yr. The  resonant argument 
$\theta$ is in libration around $\theta_0 = 76.914$ deg, with a libration 
period   $P_{\theta} \approx 19.4$ yr, and an amplitude of about $37$ degrees
(Fig.\,\ref{F7}). It should be noted that for the real solution, 
the libration amplitude may be smaller, but the libration period 
will be of the same order of magnitude, that is around 20 years.
The observation of the system over a few additional years may
then provide an estimate of the libration amplitude and thus 
a strong constraint on the parameters of the system.

\subsection{Secular evolution}

Using the S5 stable orbital parameters (Table\,\ref{T5}) we 
have first integrated our
system over a few thousand years (Fig.\,\ref{F7}). Unlike
results plotted in Fig.\,\ref{F5} for unstable systems, we now observe
a regular variation of the eccentricity of both planets.

Because of the strong gravitational interactions with the inner
planet, the outer planet still shows large variations in its
orbital parameters.  The eccentricity can range from less than 0.1 to
about 0.45, while the semi-major axes varies between 2.3 and almost
3\,AU.  As a result, the minimum distance between the two planets'
orbits is only 0.4\,AU.  However, because of the  5/1
mean motion resonance trapping, the two planets never come closer than about
1.1\,AU.  We also observe rapid secular variations of the orbital
parameters, mostly driven by the rapid secular frequency 
$g_2$, with a period $P_{g_2} \approx 339$ yr.   These secular variations of 
the orbital elements are much faster than in our Solar System, 
and should make possible their direct observation.

The S5 orbital parameters (determined using the  global view of the 
system dynamics given by Figure~\ref{F6}) 
allow us to obtain an orbital evolution of the system 
that is much  more satisfactory than the one
obtained by a direct orbital fit (section~\ref{orbsolutions}, solutions 
S3 and S4), as the system now
remains stable within five thousand years (Figs.\,\ref{F5}
and~\ref{F7}).  
Although from the previous stability analysis (section \ref{sec4.2}) 
we know that the stability of the orbit is granted for a much 
longer time interval than the few thousand years of the 
orbital integration, we have also directly tested
the stability of the system S5  over 5 Gyr. 
The results displayed in Figure~\ref{F8} show 
that indeed, the orbital elements  evolve in a
regular way, and  remain relatively stable over the age of the central star.

\begin{figure}[t!]
   \begin{tabular}{c}
    \includegraphics*[width=8.5cm]{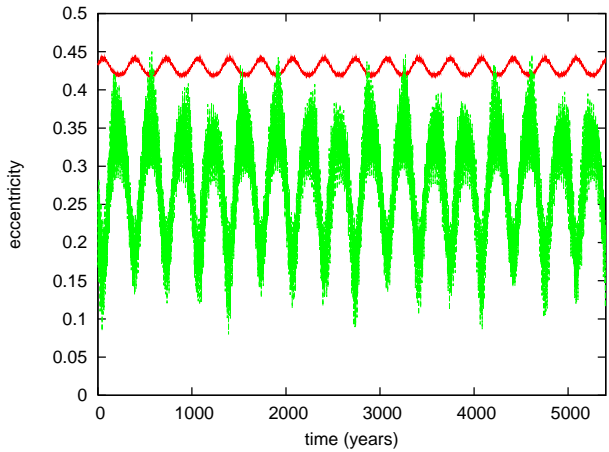}
   \end{tabular}
  \caption{Dynamical evolution of the eccentricity with
    the orbital solution S5 (Table\,\ref{T5}). As expected, the
    eccentricity presents regular variations that contrasts to the
    irregular behavior of the orbital solutions presented in
    Fig.\,\ref{F5}. Due to the strong gravitational interactions, the
    secular variations of the eccentricity are rapid, 
    and mostly driven by the secular frequency $g_2$, 
    with period  $P_{g_2} \approx 339$ yr.
    \llabel{F8}}
\end{figure}

\begin{figure}[t!]
   \begin{tabular}{c}
    \includegraphics*[width=8.5cm]{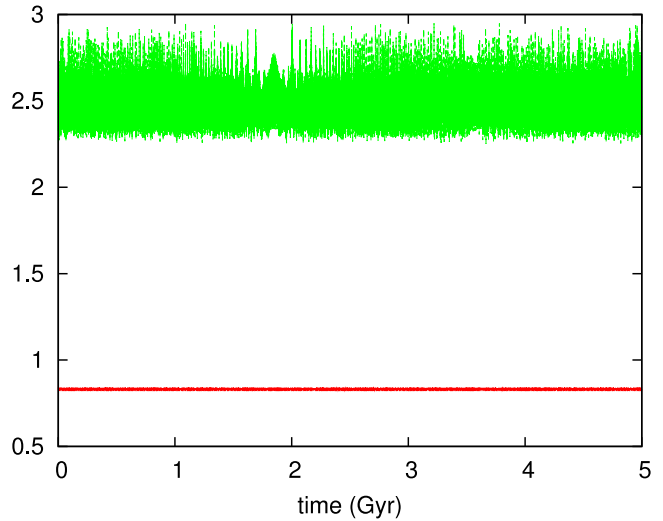} \\
    \includegraphics*[width=8.5cm]{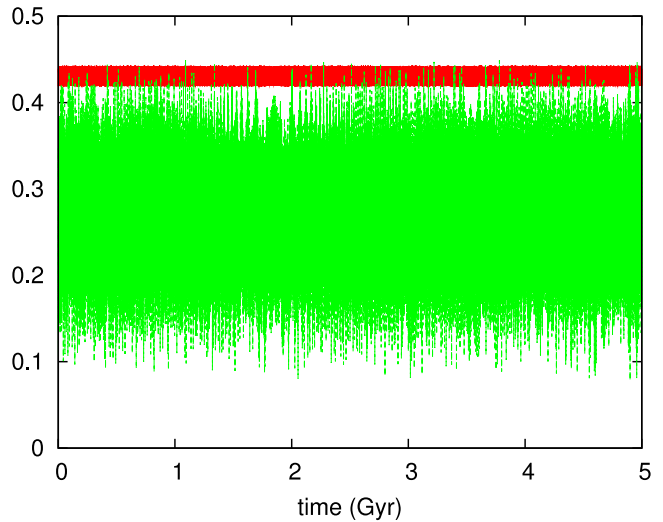}
   \end{tabular}
  \caption{Long term evolution of the semi-major axes and the eccentricity for
    both planets with the orbital solution S5
    (Table\,\ref{T5}). The system remained stable during 5 billion
    years. The small variation in the semi major axes and eccentricity
    around 2 Gyr is probably due to some very slow diffusion
    through a small resonance.
    \llabel{F9}}
\end{figure}


\section{Discussion and conclusion}

In this paper we report the presence of a second planet orbiting the
{\small HD}\,202206 star, whose orbital parameters are quite
unexpected.  This system was first described as a star orbited by a
massive planet or a light brown dwarf (Udry et al., 2002,
Paper\,I).  The first {\small CORALIE} measurements already suggested
the presence of a second, longer period companion, but it was thought
to be a very distant stellar companion.  The existence of a second,
much less massive body at only $\smaCi$\,AU, was
never observed and troubles our understanding of the hierarchy of
planetary systems.  Two other multiple planetary systems were
discovered with orbital periods identical to this one: {\small
  HD}\,12661 (264 and 1445 days) and {\small HD}\,169830 (226 and 2102
days).  However, the mass ratio of the two planets differ in both
cases by less than a factor of two, while for {\small HD}\,202206 this
ratio is almost ten.  Mazeh and Zucker (2003) suggested that a
possible correlation between mass ratio and period ratio in multiple
planetary systems may exist.  Using the multiple planetary systems
discovered to the date (including Jupiter and Saturn), they found
that, except for the 2/1 resonant systems, the correlation between the
logarithms of the two ratios was 0.9498. In order to keep this result,
the consideration of the present planetary system shows that 
mean motion resonances other than the 2/1 should probably also 
be excluded from the correlated systems.

These observations raise the question of how this system
was formed, bringing additional constraints to the existent theories.
Supposing that the inner body is effectively a brown dwarf, then the
new found planet will be an example of a planet in a binary, formed in
the circumbinary protoplanetary disk. This assumption seems to be a real
possibility, since recent numerical simulations show that a planet formed in a
circumbinary disk can migrate inward until it is captured in resonance (Nelson,
2003). Inversely, we can suppose that both companions were formed in the
accretion disk of the star, with the result that the inner planet is not a brown
dwarf.  This leads to the re-definition of the brown-dwarf limits and requires
that the initial disk around {\small HD}\,202206 was much more massive than we
would usually think.

Dynamically the system is very interesting and promising. The gravitational
interactions between the two planets are strong, but stability is possible due
to the presence of a 5/1 mean motion resonance with a libration period of about
20 years. This is the first observation of such an orbital configuration that may
have been reached through the dissipative process of planet migration 
during the early stages of the system evolution.  

The strong gravitational interactions among the planets 
may also allow us to correctly
model their effect in the nearby future.  With the current
precision of {\small CORALIE}, fixed at about 8\,m/s for {\small
  HD}\,202206, we are presently close to detecting the trace of the
planet-planet interactions in data. This will be reached even sooner with
the higher precision measurements presently obtained with the {\small
ESO HARPS} spectrograph at a $\sim$\,1\,m/s level (Mayor et al., 2003). 
The planet-planet interaction signature may provide
important information on the inclination of the orbital planes
and allow us to determine the mass values of both planets.

\paragraph{Acknowledgments} We are grateful to the staff from the Geneva
Observatory which maintains the 1.2-m Euler Swiss telescope and
the {\small CORALIE} echelle spectrograph at La Silla. This work was supported
by Geneva University, the Swiss NSF (FNRS) and from PNP-CNRS. A.C. also benefited
from the support of the HPRN-CT-2002-00308 European training network
programme.

{
\makeatletter \renewcommand\@biblabel[1]{} \makeatother   

}

\end{document}